# Fractal Analysis Based on Hierarchical Scaling in Complex Systems

Yanguang Chen

(Department of Geography, College of Urban and Environmental Sciences, Peking University, 100871, Beijing, China. Email: chenyg@pku.edu.cn)

**Abstract:** A fractal is in essence a hierarchy with cascade structure, which can be described with a set of exponential functions. From these exponential functions, a set of power laws indicative of scaling can be derived. Hierarchy structure and spatial network proved to be associated with one another. This paper is devoted to exploring the theory of fractal analysis of complex systems by means of hierarchical scaling. Two research methods are utilized to make this study, including *logic analysis method* and *empirical analysis method*. The main results are as follows. First, a fractal system such as Cantor set is described from the hierarchical angle of view; based on hierarchical structure, three approaches are proposed to estimate fractal dimension. Second, the hierarchical scaling can be generalized to describe multifractals, fractal complementary sets, and self-similar curve such as logarithmic spiral. Third, complex systems such as urban systems are demonstrated to be a self-similar hierarchy. The human settlements in Germany and the population of different languages in the world are taken as two examples to make empirical analyses. This study may be revealing for associating fractal analysis with other types of scaling analysis of complex systems, and spatial optimization theory may be developed in future by combining the theories of fractals, allometry, and hierarchy.

# 1. Introduction

Recent years, the hierarchical systems with cascade structure have attracted attention of many scientists. Representing a form of organization of complex systems, hierarchy is frequently observed within the natural world and in social institutions (Pumain, 2006). A fractal can be treated as a self-similar hierarchy because a fractal object bears many levels which are systematic arranged according to scaling laws (Batty and Longley, 1994; Chen, 2008; Frankhauser, 1998; Mandelbrot, 1983). Fractal phenomena can be described with power laws, and a power law can be decomposed into two exponential laws by means of hierarchical structure. Generally speaking, it is difficult to solve an equation based on power laws or spatial network, but it is easy to deal with the problem based on exponential models or hierarchies. Using self-similar hierarchy, we can transform fractal scaling into a hierarchical scaling with characteristic scales, thus many complex problems can be solved in a simple way. If we explore fractal systems such as a system of cities by means of hierarchy, we can use a pair of exponential laws to replace a power law, and the analytical process can be significantly simplified (Chen, 2012; Chen and Zhou, 2003). A fractal is a special case of hierarchical scaling. Hierarchy suggests a new way for understanding fractal organization and exploring complex systems.

In scientific research, three factors increase the difficulty of mathematical modeling, that is *spatial dimension*, *time lag* (response delay), and *interaction*. Economics is relatively simple because economists don't usually consider much the spatial dimension in economic systems (Waldrop, 1992). However, all the difficult problems related to mathematical modeling, especially the spatial dimension, are encountered with by geographers. If the spatial dimension is avoided, geography is not yet real geography. Geographers often study spatial structure by means of hierarchy. A discovery is that hierarchy and network structure represent two different sides of the same coin (Batty and Longley, 1994). Two typical hierarchy theories are developed in human geography. One is central place theory (Christaller, 1933/1966; Lösch, 1954), and the other, rank-size distributions (Carroll, 1982; Semboloni, 2008). The two theories are related to fractal ideas (Arlinghaus, 1985; Arlinghaus and Arlinghaus, 1989; Batty and Longley, 1994; Chen, 2011; Chen, 2014). Fractal theory, scaling concepts, and the related methods become more and more important in geographical analysis such as urban studies (Chen, 2008). As Batty (2008) once observed, "an integrated theory of how cities evolve, linking urban economics and transportation behavior to developments in network science, allometric growth, and fractal geometry, is being slowly developed." In fact, fractals, allometry, and complex network can be associated with one another in virtue of hierarchical scaling.

Hierarchical scaling suggests a new perspective to examine the simple rules hiding behind the complex systems. Many types of physical and social phenomena satisfy the well-known rank-size distribution and thus follow Zipf's law (Carroll, 1982; Chen and Zhou, 2003; Jiang and Jia, 2011). Today, Zipf's law has been used to describe the discrete the power law probability distributions in various natural and human systems (Bak, 1996; Chen, 2008). However, despite a large amount of research, the underlying rationale of the Zipf distribution is not yet very clear. On the other hand, many types of data associated with Zipf's law in the physical and social sciences can be arranged in good order to form a hierarchy with cascade structure. There are lots of evidences showing that the Zipf distribution is inherently related to the self-similar hierarchical structure, but the profound mystery has not yet to be unraveled for our understanding natural laws. The Zipf distribution is associated with fractal structure and bears an analogy with the $1/f$ fluctuation (Chen, 2012). Fractals, $1/f$ noise, and the Zipf distribution represent the observation of the ubiquitous empirical patterns in nature (Bak, 1996). This article provides scientists with a new way of looking at the relations between these ubiquitous empirical patterns and the complex evolution processes in physical and social systems, and thus to understand how nature works.

A scientific research actually includes two elements of methodology, that is, description and understanding. Science should proceed first by describing how a system works and then by understanding why (Gordon, 2005). The description process is by means of mathematics and measurement, while the understanding process is by means of observation, experience, or even artificially constructed experiments (Henry, 2002). This work is devoted to exploring fractal modeling and spatial analysis based on hierarchy with cascade structure. First of all, we try to describe and understand hierarchy itself; later, we try to use hierarchical scaling to describe and understand complex systems. Two research methods are utilized in this works. One is *logic analysis method*, including induction method and deduction method, and the other is *empirical analysis method*, fitting the mathematical models to observational data. The induction method is based on various regular fractals such as Cantor set, Koch snowflake curve, Vicsek box, and Sierpinski gasket, while the deduction method is mainly based on mathematical derivation. As for empirical analysis, systems of cities and population size distribution of languages can be taken as examples. Anyway, the success of natural sciences lies in their great emphasis on the interplay between quantifiable data and models (Louf and Barthelemy, 2014). The rest parts are organized as follows. In Section 2, a set of hierarchical models of fractals, including monofractals and multifractals, are presented. In Section 3, case studies are made by means of Germany systems of human settlements and world population size of different languages. In Section 4, several questions are discussed, and the hierarchical scaling modeling is generalized. Finally, the discussion is concluded by summarizing the main points of this work.

## 2. Models

### 2.1 Three approaches to estimating fractal dimension

A regular fractal is a typical hierarchy with cascade structure. Let's take the well-known Cantor set as an example to show how to describe the hierarchical structure and how to calculate its fractal dimension (Figure 1). We can use two measurements, the length ($L$) and number ($N$) of fractal copies in the $m$th class, to characterize the self-similar hierarchy. Thus, we have two exponential functions such as

$$N_m = N_1 r_n^{m-1} = \frac{N_1}{r_n} e^{(\ln r_n)m} = N_0 e^{\omega m}, \tag{1}$$

$$L_m = L_1 r_l^{1-m} = L_1 r_l e^{-(\ln r_l)m} = L_0 e^{-\psi m}, \tag{2}$$

where $m$ denotes the ordinal number of class ($m$=1, 2, …), $N_m$ is the number of the fractal copies of a given length, $L_m$ is the length of the fractal copies in the $m$th class, $N_1$ and $L_1$ are the number and length of the initiator ($N_1$=1), respectively, $r_n$ and $r_l$ are the **number ratio** and **length ratio** of fractal copies, $N_0 = N_1/r_n$, $L_0 = L_1 r_l$, $\omega = \ln(r_n)$, $\psi = \ln(r_l)$. From equations (1) and (2) it follows the common ratios of number and length, that is

$$r_n = \frac{N_1 r_n^m}{N_1 r_n^{m-1}} = \frac{N_{m+1}}{N_m}, \tag{3}$$

$$r_l = \frac{L_1 r_l^{1-m}}{L_1 r_l^{-m}} = \frac{L_m}{L_{m+1}}. \tag{4}$$

According to the definitions of $\omega$ and $\psi$, the logarithms of equations (3) and (4) are

$$\omega = \ln(r_n) = \ln(\frac{N_{m+1}}{N_m}), \tag{5}$$

$$\psi = \ln(r_l) = \ln(\frac{L_m}{L_{m+1}}). \tag{6}$$

From equations (1) and (2), we can derive a power law in the form

$$N_m = kL_m^{-D}, \tag{7}$$

in which $k=N_1L_1^D$ is the proportionality coefficient, and $D=\ln(r_n)/\ln(r_l)$ is the fractal dimension of the Cantor set ($k=1$). Thus, three formulae of fractal dimension estimation can be obtained. Based on the power law, the fractal dimension can be expressed as

$$D = -\frac{\ln(N_m)}{\ln(L_m)}. \tag{8}$$

Based on the exponential models, the fractal dimension is

$$D = \frac{\omega}{\psi}. \tag{9}$$

Based on the common ratios, the fractal dimension is

$$D = \frac{\ln r_n}{\ln r_l}. \tag{10}$$

In theory, equations (8), (9), and (10) are equivalent to one another. Actually, by recurrence, equation (7) can be rewritten as $N_{m+1}=L_{m+1}^{-D}$, and thus we have

$$\frac{N_{m+1}}{N_m} = (\frac{L_{m+1}}{L_m})^{-D}, \tag{11}$$

Taking logarithms on both sides of equation (11) yields

$$D = -\frac{\ln(N_{m+1}/N_m)}{\ln(L_{m+1}/L_m)} = -\frac{\ln(N_m)}{\ln(L_m)} = \frac{\ln r_n}{\ln r_l} = \frac{\omega}{\psi}. \tag{12}$$

For the Cantor set, $N_m=2^{m-1}$, $L_m=1/3^{m-1}$, $r_n=N_{m+1}/N_m=2$, $r_l=L_m/L_{m+1}=3$, $\omega=\ln(2)$, $\psi=\ln(3)$, thus we have

$$D = \frac{\ln(2)}{\ln(3)} \approx 0.631.$$

This suggests that, for the regular fractal hierarchy, three approaches lead to the same result. The fractal dimension can be computed by using exponential functions, power function, or common ratios, and all these values are equal to one another. However, in practice, there are subtle differences between the results from different approaches because of random noise in observational data. Of course, the differences are not significant and thus can be negligible.

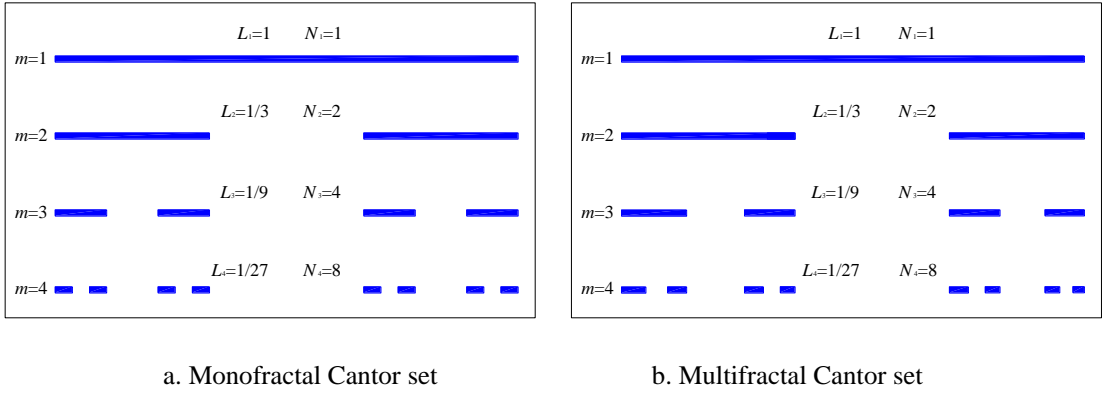

        a. Monofractal Cantor set          b. Multifractal Cantor set

**Figure 1 The Cantor set as a self-similar hierarchy with cascade structure (the first four classes)**

[**Note**: An analogy can be drawn between the Cantor sets and self-hierarchies of human settlements. For the multifractal Cantor set, the length of each level is average value.]

The mathematical description and fractal dimension calculation of the Cantor set can be generalized to other regular fractals such as Koch snowflake and Sierpinski gasket or even to the route from bifurcation to chaos. As a simple fractal, the Cantor set fails to follow the rank-size law. However, if we substitute the multifractal structure for monofractal structure, the multi-scaling Cantor set will comply with the rank-size rule empirically.

**2.2 Multifractal characterization of hierarchies**

Monofractals (unifractals) represent the scale-free systems of homogeneity, while multifractals represents the scale-free systems of heterogeneity. In fact, as Stanley and Meakin (1988) pointed out, "multifractal scaling provides a quantitative description of a broad range of heterogeneous phenomena." In geography, multifractal geometry is a powerful tool for describing spatial heterogeneity. A multifractal hierarchy of Cantor set can be organized as follows. At the first level, the initiator is still a straight line segment of unit length, i.e., $S_1=L_1=1$. At the second level, the generator includes two straight line segments of different lengths. The length of one segment is $a$, and the other's length is $b$. Let $a=3/8$, $b=2/3-a=7/24$. The summation of the two line segments' length is 2/3, that is, $S_2=a+b=2/3$, and the average length of the two segments is $L_2=S_2/2=1/3$. At the third level, there are four line segments, and the lengths are $a^2$, $ab$, $ba$, and $b^2$, respectively.

The total length of the four line segments is 4/9, namely, $S_3=(a+b)^2=(2/3)^2$, and the average length is $L_3= S_3/4=1/3^2$. Generally speaking, the $m$th level consists of $2^{m-1}$ line segments with lengths of $a^{m-1}$, $a^{m-2}b$, $a^{m-3}b^2$, …, $a^2b^{m-3}$, $ab^{m-2}$, and $b^{m-1}$, respectively. The length summation is $S_m=(a+b)^{m-1} = (2/3)^{m-1}$, so the average length is

$$L_m = \frac{(a+b)^{m-1}}{N_m} = 3^{1-m}, \tag{13}$$

where

$$N_m = 2^{m-1}. \tag{14}$$

From equations (13) and (14) it follows a scaling relation as below:

$$N_m = L_m^{-\ln(2)/\ln(3)} = \mu L_m^{-D_0}, \tag{15}$$

which is identical in form to equation (7), and the capacity dimension $D_0=\ln(2)/\ln(3)\approx 0.631$ is equal to the fractal dimension of the monofractal Cantor set (Figure 1).

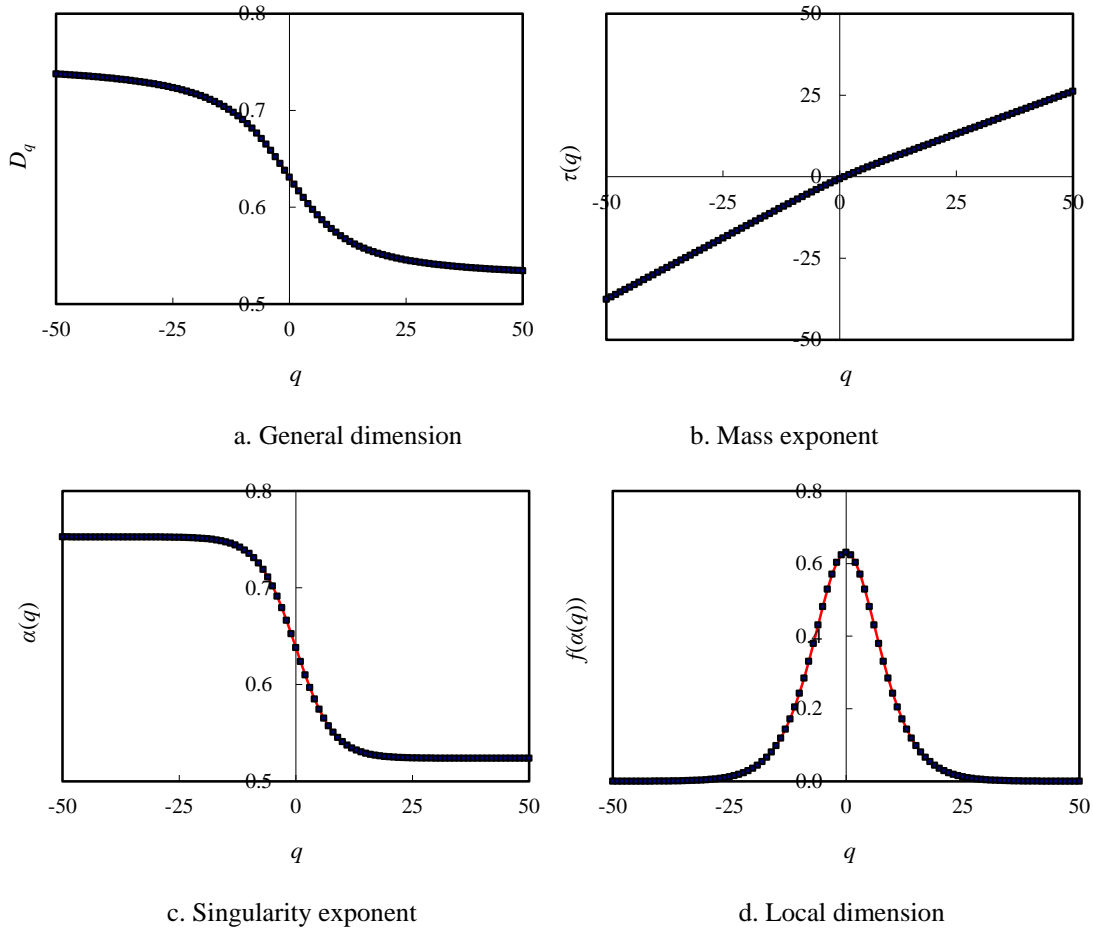

a. General dimension  b. Mass exponent

c. Singularity exponent  d. Local dimension

**Figure 2 The dimension spectrums of multifractals Cantor set and the curves of related parameters ($p=9/16$)**

Two sets of parameters are always employed to characterize a multifractal system. One is the set of *global* parameters, and the other, the set of *local* parameters. The global parameters include the generalized correlation dimension and the mass exponent; the local parameters comprise the Lipschitz-Hölder exponent and the fractal dimension of the set supporting this exponent. For the two-scale Cantor set, the mass exponent is

$$\tau(q) = -\frac{\ln[p^q + (1-p)^q]}{\ln(3)}, \qquad (16)$$

where $q$ denotes the moment order ($-\infty < q < \infty$), $\tau(q)$ refers to the mass exponent, and $p$ is a probability measurement. Take the derivative of equation (16) with respect to $q$ yields the Lipschitz-Hölder exponent of singularity in the form

$$\alpha(q) = \frac{d\tau(q)}{dq} = -\frac{1}{\ln(3)} \frac{p^q \ln p + (1-p)^q \ln(1-p)}{p^q + (1-p)^q}, \qquad (17)$$

in which $\alpha(q)$ refers to the singularity exponent. Utilizing the Legendre transform, we can derive the fractal dimension of the subsets supporting the exponent of singularity such as

$$f(\alpha) = q\alpha(q) - \tau(q) = \frac{1}{\ln(3)}[\ln[p^q + (1-p)^q] - \frac{p^q \ln p^q + (1-p)^q \ln(1-p)^q}{p^q + (1-p)^q}], \qquad (18)$$

where $f(\alpha)$ denotes the local dimension of the multifractal set. Furthermore, the general fractal dimension spectrum can be given in the following form

$$D_q = \begin{cases} -\dfrac{p \ln p + (1-p)\ln(1-p)}{\ln(3)}, & q = 1 \\ \dfrac{\tau(q)}{q-1}, & q \neq 1 \end{cases}, \qquad (19)$$

where $D_q$ denotes the generalized correlation dimension. If the order moment $q \neq 1$, the general dimension can also be expressed as

$$D_q = \frac{1}{q-1}[q\alpha(q) - f(\alpha)]. \qquad (20)$$

Using the above equations, we can describe multifractal Cantor set. For example, if the length of one line segment in the generator is $a=3/8$ as assumed, then the length of another line segment is $b=7/24$. Accordingly, the probability measures are $p=a/(2/3)=9/16$ and $1-p=7/16$. By means of these formula, the multifractal dimension spectra and the related curves can be displayed in Figures 2 and 3. The capacity dimension is $D_0 \approx 0.631$, the information dimension is $D_1 \approx 0.624$, and the correlation dimension is $D_2 \approx 0.617$. Substituting ln(2) for ln(3) in the equations shown above, we can use the multifractal models of Cantor set to describe multi-scaling rank-size distribution of cities (Chen, 2012; Chen and Zhou, 2004).

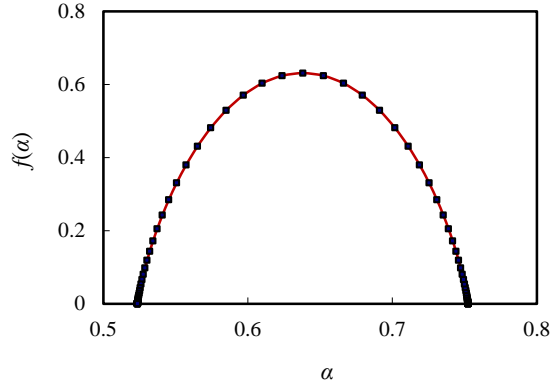

**Figure 3 The *f(α)* curves of the local dimension v.s. the singularity exponent**

## 2.3 Hierarchical scaling in social systems

Fractal hierarchical scaling can be generalized to model general hierarchical systems with cascade structure. Suppose dividing the elements (e.g. cities) in a large-scale system (e.g. a regional system) into *M* levels in the top-down order. We can describe the hierarchical structure using a set of exponential functions as follows

$$N_m = N_1 r_f^{m-1}, \tag{21}$$

$$P_m = P_1 r_p^{1-m}, \tag{22}$$

$$A_m = A_1 r_a^{1-m}. \tag{23}$$

where *m* denotes the top-down order (*m*=1, 2, … *M*), $N_m$ represents the element number in a given order, clearly, $r_n = N_{m+1}/N_m$ is actually the **number ratio**, $N_1$ is the number of the top-order elements--Generally speaking, we have $N_1$=1; $P_m$ represents the mean size of order *m*, $r_p = P_m/P_{m+1}$ is element **size ratio** of adjacent levels, $P_1$ is the mean size of the first-order elements, i.e. the largest ones; $A_m$ is the average area of order *m*, $r_a = A_m/A_{m+1}$ is **area ratio**, and $A_1$ is the area of the first order. Rearranging equation (22) yields $r_p^{m-1} = P_1/P_m$, then taking logarithm to the base $r_n$ of this equation and substituting the result into equation (21) yields a power function as

$$N_m = \mu P_m^{-D}, \tag{24}$$

where $\mu = N_1 P_1^D$, $D = \ln(r_n)/\ln(r_p)$. Equation (24) is hereafter referred to as the "size-number law", and $D$ proved to be the fractal paradimension of self-similar hierarchies measured by city population size. Similarly, from equations (21) and (23) it follows

$$N_m = \eta A_m^{-d}, \tag{25}$$

in which $\eta = N_1 A_1^d$, $d = \ln(r_n)/\ln(r_a)$. equation (25) is what is "area-number law", and $d$ is the fractal paradimension of self-similar hierarchies measured by urban area. Finally, we can derive the hierarchical allometric scaling relationships between area and size from equations (22) and (23), or from equations (24) and (25), and the result is

$$A_m = a P_m^b, \tag{26}$$

where $a = A_1 P_1^{-b}$, $b = \ln r_a / \ln r_p$. This is just the generalized allometric growth law on the area-size relations. Further a three-parameter Zipf-type model on size distribution can be derived from equations (1) and (2) such as

$$P_k = C(k + \varsigma)^{-\alpha}, \tag{27}$$

where $k$ is the rank among all elements in a given system in decreasing order of size, $P_k$ is the size of the $k$th element. As the parameters, we have the constant of proportionality $C = P_1[r_n/(r_n-1)]^{1/D}$, the small parameter $\varsigma = 1/(r_n-1)$, and the power $\alpha = 1/D = \ln r_p / \ln r_f$. Where $P_1$ is the size of the largest element, $q$ proved to be the reciprocal of the fractal dimension $D$ of city-size distribution or urban hierarchies, i.e. $\alpha = 1/D$ (Chen and Zhou, 2003). By analogy, we can derive a three-parameter Zipf-type model on area distribution from equations (1) and (3), that is

$$A_k = G(k + \zeta)^{-\beta}, \tag{28}$$

where $G$, $\zeta$, and $\beta$ are parameters. In theory, $\beta = 1/d$. From equation (27) and (28) it follows

$$(\frac{P_k}{C})^{-1/\alpha} - \varsigma = (\frac{A_k}{G})^{-1/\beta} - \zeta, \tag{29}$$

which suggests an approximate allometric relation. If $\zeta = \varsigma$, then we can derive a cross-sectional allometry relation between size and area from equation (27) and (28) as below

$$A_k = a P_k^b, \tag{30}$$

where $a=A_1P_1^{-b}$, $b=\beta/\alpha$. Equation (30) is mathematically equivalent to equation (26), that is, the rank-size allometric scaling is equivalent to hierarchical allometric scaling in theory. Further, if $\zeta=\varsigma=0$, then equations (27) and (28) will be reduced to the common two-parameter Zipf's models (Chen, 2008). The fractal models (principal scaling laws), allometric model (the law of allometric growth), and rank-size distribution model (Zipf's law), are three basic scaling laws of hierarchical systems such as cities, and all the scaling relations can be derived from the hierarchical models expressed by exponential functions.

## 3 Empirical analysis

### 3.1 A case of Germany "natural cities"

#### 3.1.1 Material and data

First of all, the hierarchy of German cities is employed to illustrate hierarchical scaling method. Recently, Bin Jiang and his coworkers have proposed a concept of "natural city" and developed a novel approach to measure objective city sizes based on street nodes or blocks and thus urban boundaries can be naturally identified (Jiang and Liu, 2012; Jiang and Jia, 2011). The street nodes are defined as street intersections and ends, while the naturally defined urban boundaries constitute the region of what is called *natural cities*. The street nodes are significantly correlated with population of cities as well as city areal extents. The city data are extracted from massive volunteered geographic information OpenStreetMap databases through some data-intensive computing processes and three data sets on European cities, including the cities of **France**, **Germany**, and the **United Kingdom** (UK), have been obtained. Among all these data sets, the set for German is the largest one, which encompasses the 5160 largest natural cities. Therefore, German cities are taken as an example to make empirical analysis. In the processing of data, the area variable is divided by 10000 for comparability.

#### 3.1.2 Method and results

The analytical method is based on the theoretical models shown above. For the natural cities, the population size measurement ($P$) should be replaced by the amount of blocks in the physical areal extent ($A$), which can be treated as a new size measurement of cities. It is easy to use

German cities to construct a hierarchy to illustrate equivalence relation between the rank-size law and the hierarchical scaling. Empirically, the largest 5160 cities and towns follow the rank-size rule and we have

$$\hat{P}_k = 160175.044 k^{-1.051},$$

where $k$ is the rank of natural cities, and $P_k$ the city size defined with urban blocks in objective boundaries. The symbol "^" implies "estimated value", "calculated value", and "predicted value". The goodness of fit is about $R^2=0.993$, and the scaling exponent is around $q=1.051$ (Figure 4).

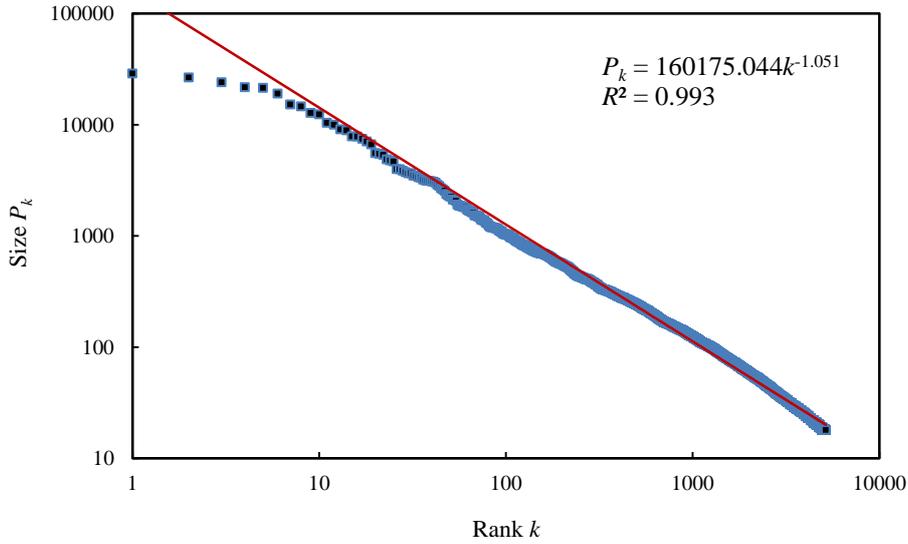

**Figure 4 The rank-size pattern of Germany cities by blocks within physical extent (2010)**

The models of fractals and allometry can be built for German hierarchies of cities as follows. Taking number ratio $r_f=2$, we can group the cities into different classes according to the $2^n$ rule (Chen and Zhou, 2003; Davis, 1978). The results, including city number ($N_m$), total amount of urban blocks ($S_m$), average size by blocks ($P_m$), total area ($T_m$), and average area ($A_m$), in each class, are listed in Table 1. In a hierarchy, two classes, i.e., top class and bottom class, are always special and can be considered to be exceptional values (Figure 5). In fact, the power law relations always break down if the scale is too large or too small (Bak, 1996). Thus a scaling range can be found in a log-log plot of fractal analysis on cities (Chen, 2008). Two hierarchical scaling relations can be testified by the least squares calculation. For common ratio $r_n=2$, the hierarchical scaling relation between city size and number is

$$\hat{N}_m = 91161.315 P_m^{-1.025}.$$

The goodness of fit is about $R^2=0.996$, and the fractal dimension of the self-similar hierarchy is $D\approx1.025$. The average size ratio within the scaling range is about $r_p=1.942$, which is very close to $r_n=2$. Thus another fractal dimension estimation is $D=\ln(r_n)/\ln(r_p) \approx1.045$. The average size follows the exponential law, that is, $P_m=133869.061*\exp(-0.674m)$. Thus the third fractal dimension estimation is $D=\omega/\psi\approx0.693/0.674\approx1.028$. All these results are based on scaling range rather than the whole classes. Similarly, the relation between urban area and city number is as below

$$\hat{N}_m = 162295.381 A_m^{-1.060}.$$

The goodness of fit is about $R^2=0.998$, and the fractal dimension of the self-similar hierarchy is $d\approx1.060$. The average size ratio within the scaling range is about $r_p=1.927$. So another fractal dimension estimation is $d=\ln(r_n)/\ln(r_a) \approx1.056$. The average area follows the exponential law, that is, $A_m=157737.532*\exp(-0.653m)$. Thereby, the third fractal dimension estimation is $d=\omega/\psi\approx 0.693/0.653 \approx1.061$. Further, by means of the datasets of urban size and area, an allometric scaling model can be built as follows

$$\hat{A}_m = 1.722 P_m^{0.967}.$$

The goodness of fit is around $R^2=0.999$, and the scaling exponent $b\approx0.967$ (Figure 6). Another estimation of the allometric exponent is $b\approx1.025/1.060\approx0.967$. The two results are close to one another. The natural cities of Germany lend further support to the equivalent relationship between the rank-size distribution and the self-similar hierarchy.

Table 1. The size and number for the hierarchy of German's cities based on the $2^n$ principle

| $m$ | Total block ($S_m$) | Total area ($T_m$) | Average size ($P_m$) | Average area ($A_m$) | Number ($N_m$) |
|---|---|---|---|---|---|
| 1 | 28866 | 402657796.2 | 28866.0 | 402657796.2 | 1 |
| 2 | 50709 | 731271674.1 | 25354.5 | 365635837.1 | 2 |
| 3 | 77576 | 1030661786.8 | 19394.0 | 257665446.7 | 4 |
| 4 | 86071 | 973558025.6 | 10758.9 | 121694753.2 | 8 |
| 5 | 82700 | 999267240.9 | 5168.8 | 62454202.6 | 16 |
| 6 | 80912 | 940916731.4 | 2528.5 | 29403647.9 | 32 |
| 7 | 72397 | 986813213.3 | 1131.2 | 15418956.5 | 64 |
| 8 | 75375 | 1070810188.5 | 588.9 | 8365704.6 | 128 |
| 9 | 79299 | 1165806475.4 | 309.8 | 4553931.5 | 256 |
| 10 | 84327 | 1271861134.1 | 164.7 | 2484103.8 | 512 |

| 11 | 84599 | 1310854103.8 | 82.6 | 1280131.0 | 1024 |
| 12 | 75214 | 1138100595.1 | 36.7 | 555713.2 | 2048 |
| 13 | 21820 | 197476690.8 | 20.5 | 185424.1 | 1065* |

**Source**: The original data come from Jiang (http://arxiv.org/find/all/). ***Note**: The last class of each hierarchy is a *lame-duck class* termed by Davis (1978).

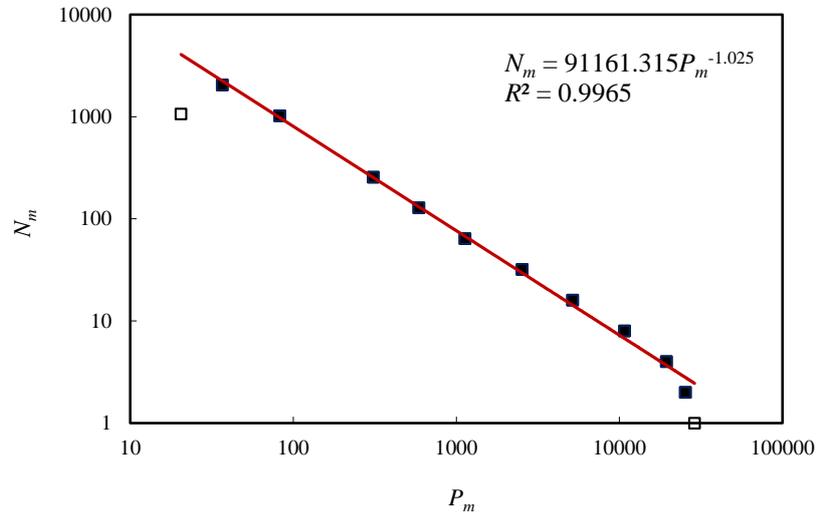

a. Size-number scaling

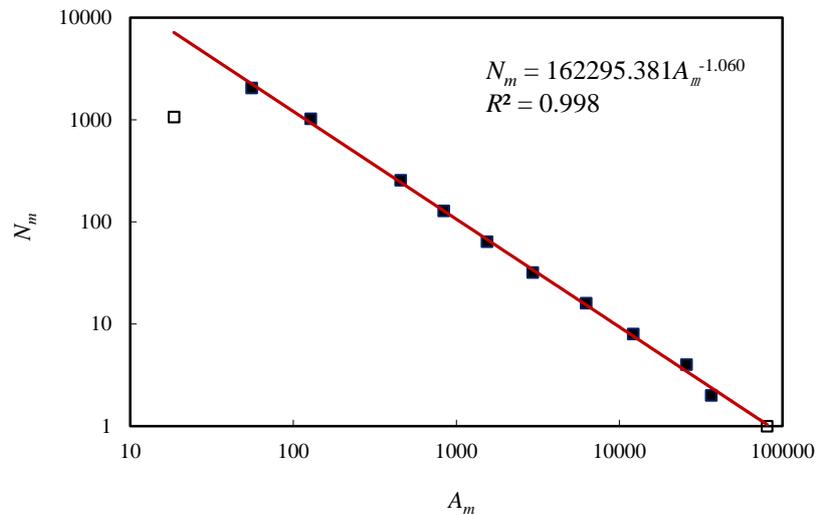

b. Area-number scaling

**Figure 5 The scaling relations between city numbers and average sizes/area in the hierarchies of German cities by the blocks in physical extent in 2010**

[**Note**: The hollow squares represent the outliers, while the solid squares form a scaling range. For simplicity, the urban area is rescaled by dividing it using 10,000.]

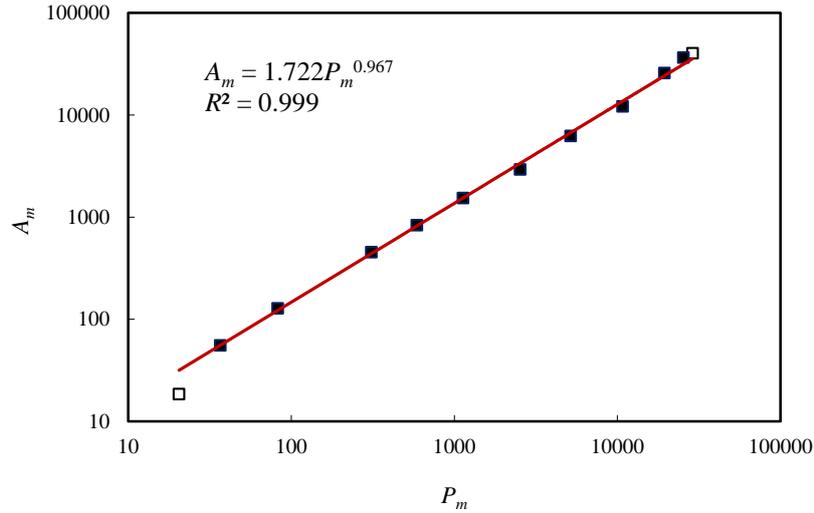

**Figure 6 The allometric scaling relations between average population sizes and urban areas in the hierarchies of Germany cities in 2010**

[**Note**: The hollow square represents the outlier, while the solid squares form a scaling range. For comparability, the area measure is divided by 10,000.]

## 3.2 A case of language hierarchy in the world

The hierarchical scaling can be used to model the rank-size distribution of languages by population. Where population size is concerned, there are 107 top languages in the world such as Chinese, English, and Spanish. In data processing, the population size is rescaled by dividing it with 1,000,000 for simplicity. Gleich *et al* (2000) gave a list of the 15 languages by number of native speakers (Table 2). The rank-size model of the 107 languages is as below:

$$\hat{P}_k = 1092.160 k^{-1.053},$$

where $k$ refers to rank, and $P_k$, to the population speaking the language ranked $k$, the goodness of fit is about $R^2=0.986$ (Figure 7). The fractal dimension is estimated as $D\approx0.949$.

**Table 2. The self-similar hierarchy of the 15 top languages by population**

Unit: million

| Level | Number | Language and population | | | | Total population | Average population | Size ratio |
|---|---|---|---|---|---|---|---|---|
| 1 | 1 | Chinese | 885 | | | 885 | 885 | |
| 2 | 2 | English | 470 | Spanish | 332 | 802 | 401 | 2.207 |
| 3 | 4 | Bengali | 189 | Portuguese | 170 | 711 | 177.75 | 2.256 |
| | | Indic | 182 | Russian | 170 | | | |
| 4 | 8 | Japanese | 125 | Korean | 75 | 657 | 82.125 | 2.164 |

| | | German | 98 | French | 72 | | | |
|---|---|---|---|---|---|---|---|---|
| | | Wu-Chinese | 77 | Vietnamese | 68 | | | |
| | | Javanese | 76 | Telugu | 66 | | | |

**Source**: Gleich M, *et al* (2000). **Note**: If we use the lower limits of population size $s_1$=520, $s_2$=260, $s_3$=130, and $s_4$=65 to classify the languages in the table, the corresponding number of languages is $f_1$=1, $f_2$=2, $f_3$=4, and $f_4$=8, and the scaling exponent is just 1.

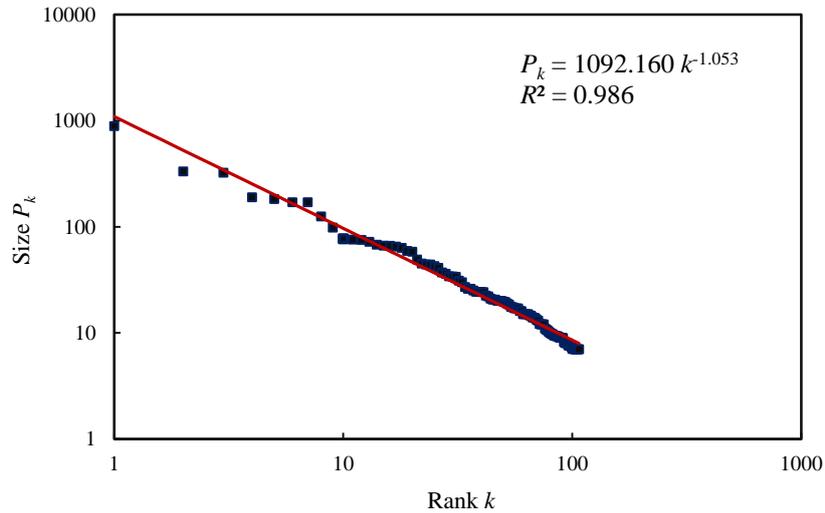

**Figure 7 The rank-size pattern of the top 107 languages by population** (Unit: million)

$P_k = 1092.160 k^{-1.053}$
$R^2 = 0.986$

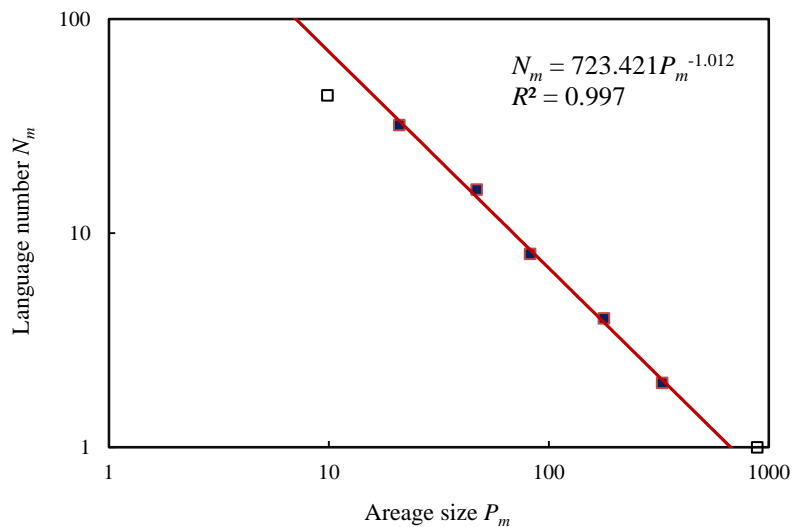

**Figure 8 The hierarchical scaling relationships between population size and number of languages**

(**Unit**: million)

[**Note:** The small circles represent top classes and the lame-duck classes, respectively. Removing the first and last classes yields a scaling range. The slope based on the scaling range indicates the fractal parameters of rank-size distributions.]

$N_m = 723.421 P_m^{-1.012}$
$R^2 = 0.997$

Using the hierarchical scaling, we can estimate the fractal dimension of the size distribution of languages in the better way. According to the $2^n$ rule, the 107 languages fall into 8 classes by size (Table 3). In the top level, 1 language, i.e., Chinese, and the total of Chinese-speaking population is 885 million; in the second level, two languages, English and Spanish, with total population 654 million.... The number ratio is defined as $r_n=2$. The corresponding size ratio is around $r_p=2.025$. Thus, the fractal dimension can be estimated as $D=\ln(r_n)/\ln(r_p)=\ln(2)/\ln(2.025)=0.983$, which is close to the reciprocal of Zipf exponent, 0.949. A regression analysis yields a hierarchical scaling relation between language number, $N_m$, and average population size, $S_m$, such as

$$\hat{N}_m = 723.421 P_m^{-1.012}.$$

The squared correlation coefficient is $R^2=0.997$, and the fractal dimension is about $D=1.012$, which is close to the above-estimated value, 0.983 (Figure 8). The results suggest that the languages by population and cities by population follow the same hierarchical scaling laws.

Table 3. The self-similar hierarchy of the 107 top languages by population size

| Level $m$ | Total population $S_m$ | Number $N_m$ | Average size $S_m$ | Size ratio $r_p$ |
|---|---|---|---|---|
| 1 | 885000000 | 1 | 885000000.0 |  |
| 2 | 654000000 | 2 | 327000000.0 | 2.706 |
| 3 | 711000000 | 4 | 177750000.0 | 1.840 |
| 4 | 656687800 | 8 | 82085975.0 | 2.165 |
| 5 | 751058000 | 16 | 46941125.0 | 1.749 |
| 6 | 668446000 | 32 | 20888937.5 | 2.247 |
| 7 | 433020412 | 44 | 9841373.0 | 2.123 |

**Note:** The source of the original data: http://www.nationmaster.com/ . The number ratio is 2. The first class is exceptional, and the last class is a lame-duck class, which is defined by Davis (1978). By the way, there is subtle difference of English population between Table 2 and Table 3, but this error does not influence the conclusions.

## 4 Questions and discussion

### 4.1 Hierarchical scaling: a universal law

A complex system is always associated with hierarchy with cascade structure, which indicates self-similarity. A self-similar hierarchy such as cities as systems and systems of cities can be described with three types of scaling laws: *fractal laws*, *allometric law*, and *Zipf's law*. These scaling laws can be expressed as equations from (24) to (30). Hierarchical scaling is a universal law in nature and human society, and it can be utilized to characterize many phenomena with different levels. Besides fractals, it can be used to depict the routes from bifurcation to chaos (Chen, 2008). In geomorphology, the hierarchical scaling has been employed to describe river systems (Horton, 1945; Rodriguez-Iturbe and Rinaldo, 2001; Schumm, 1956; Strahler, 1952); In geology and seismology, it is employed to describe the cascade structure of earthquake energy distributions (Gutenberg and Richter, 1954; Turcotte, 1997); In biology and anatomy, it is used to describe the geometrical morphology of coronary arteries in human bodies and dogs (Chen, 2015; Jiang and He,1989; Jiang and He, 1990). In urban geography, it is used to describe central place systems and self-organized network of cities (Chen, 2008; Chen and Zhou, 2003). In short, where there is a rank-size distribution, there is cascade structure; and where there is cascade structure, there is hierarchical scaling relations.

Next, hierarchical scaling is generalized to describe fractal complementary sets and quasi-fractal structure, which represent two typical cases of hierarchical description besides fractals. The basic property of fractals is self-similarity. For convenience of expression and reasoning, the concept of self-similarity point should be defined. A fractal construction starts from an initiator by way of generator. If a fractal's generator has two parts indicative of two fractal units, the fractal bears two self-similarity points; if a fractal's generator has three parts, the fractal possesses three self-similarity points, and so on. For example, Cantor set has two self-similarity points, Sierpinski gasket has three self-similarity points, Koch curve has four self-similarity points, and the box growing fractal has five self-similarity points. The number of self-similarity points is equal to the number ratio, i.e., the common ratio of fractal units at different levels. A real fractal bears at least two self-similarity points, this suggests cross similarity of a fractal besides the self-similarity. Self-similarity indicates dilation symmetry, which cross similarity implies translation symmetry. However, if and only if a system possesses more than one self-similarity point, the system can be treated as s real fractal system, and this system can be characterized by fractal geometry. A fractal bears both dilation and translation symmetry. The systems with only one self-similarity point such as logarithmic spiral can be described with hierarchical scaling. This implies that we can supplement fractal analysis by means of hierarchical scaling.

## 4.2 Hierarchies of fractal complementary sets

A fractal set and its complementary set represent two different sides of the same coin. The dimension of a fractal is always a fractional value, coming between the topological dimension and the Euclidean dimension of its embedding space. Of course, the similarity dimension is of exception and may be greater than its embedding dimension. The dimension of the corresponding complement, however, is equal to the Euclidean dimension of the embedding space. Anyway, the Lebesque measure of a fractal set is zero; in contrast, the Lebesgue measure of the fractal complement is greater than zero. Let us see the following patterns. Figure 9(a) shows the generator (i.e., the second step) of Vicsek's growing fractal set (Vicsek, 1989), which bears an analogy with urban growth; Figure 9(b) illustrates the complementary set of the fractal set (the second step). It is easy to prove that the dimension of a fractal's complement is a Euclidean dimension. If we us box-counting method to measure the complement of a fractal defined in a 2-dimension space, the extreme of the nonempty box number is

$$C_m = \lim_{m \to \infty}(r_l^{2(m-1)} - r_n^{m-1}) \to r_l^{2(m-1)}, \tag{31}$$

where $C_m$ denotes the nonempty box number for fractal complement, the rest notation is the same as those in equations (1) and (2). Thus the dimension of the fractal complement set is

$$d = -\lim_{m \to \infty} \frac{\ln C_m}{\ln L_m} = -\lim_{m \to \infty} \frac{\ln(r_l^{2(m-1)} - r_n^{m-1})}{\ln(r_l^{1-m})} \to 2, \tag{32}$$

which is equal to the Euclidean dimension of the embedding space.

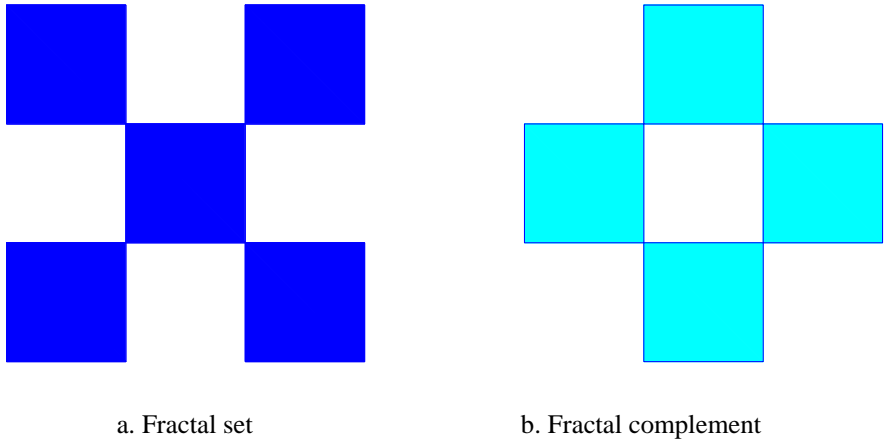

a. Fractal set     b. Fractal complement

**Figure 9 A schematic representation of fractal set and its complementary set (the 2nd step)**

However, a fractal set and its complement are of unity of opposites. A thin fractal is characterized with the fractal parameter, and the value of a fractal dimension is determined by both the fractal set and its complement. Without fractal dimension, we will know little about a fractal; without fractal complement, a fractal will degenerate to a Euclidean geometrical object. This suggests that the fractal dimension of a fractal can be inferred by its complement by means of hierarchical scaling. For example, in fractal urban studies, an urban space includes two parts: one is fractal set, and the other, fractal complement. If we define a fractal city in a 2-dimensional space, the form of urban growth can be represented by a built-up pattern, which composes of varied patches in a digital map. Further, if we define an urban region using circular area or square area, the blank space in the urban region can be treated as a fractal complement of a city. Of course, a self-organized system such as cities in the real world is more complicated than the regular fractals in the mathematical world. The differences between fractal cities and real fractals can be reflected by the models and parameters in the computational world.

A set of exponential functions and power laws can be employed to characterize the hierarchical structure of fractal complementary sets. Suppose the number of fractal units in a generator is $u$, and the corresponding number of the complementary units in the generator is $v$. For example, for Cantor set, $u=2$, $v=1$ (Figure 1); for Koch curve, $u=4$, $v=1$(Figure 10); For Sierpinski gasket, $u=3$, $v=1$(Figure 11); for Vicsek fractal, $u=5$, $v=4$ (Figure 9; Figure 12); and so on. Thus a fractal complement can be described by a pair of exponential function as below:

$$N_m = \frac{v}{u} r_n^{m-1} = v r_n^{m-2}, \quad (33)$$

$$L_m = L_1 r_l^{1-m}, \quad (34)$$

where the parameter $u=r_n$. That is, the number of fractal units in the generator is equal to the number ratio of the fractal hierarchy. Obviously, equation (33) is proportional to equation (1), while equation (34) is identical to equation (2). From equations (33) and (34) it follows

$$N_m = c L_m^{-D}, \quad (35)$$

in which $c=(v/u)L_1^D$ is the proportionality coefficient, and $D=\ln(r_n)/\ln(r_l)$ is the fractal dimension. This suggests that we can estimate the dimension value of a fractal by means of its complement. For a fractal defined in a 2-dimensional embedding space, the dimension of the complementary set is $d=2$. However, we can calculate the fractional dimension of the fractal through the scaling exponent of the complement. For instance, the exponent of the hierarchical scaling relation between scale and number in different levels of the complement of Sierpinski gasket is $D=\ln(3)/\ln(2)=1.585$, which is just the fractal dimension of Sierpinski gasket itself. The other fractal can be understood by analogy (Table 4).

Table 4. The relationships and differences between hierarchies of fractal sets and corresponding hierarchies of complementary sets (4 typical examples)

| Level | Cantor set | | | Koch curve | | |
|---|---|---|---|---|---|---|
| | Scale $L_m$ | Number $N_m$ | | Scale $L_m$ | Number $N_m$ | |
| | | Fractal | Complement | | Fractal | Complement |
| 1 | $1/3^0$ | $2^0$ | $(2^{-1})$ | $1/3^0$ | $4^0$ | $(4^{-1})$ |
| 2 | $1/3^1$ | $2^1$ | $2^0$ | $1/3^1$ | $4^1$ | $4^0$ |
| 3 | $1/3^2$ | $2^2$ | $2^1$ | $1/3^2$ | $4^2$ | $4^1$ |
| 4 | $1/3^3$ | $2^3$ | $2^2$ | $1/3^3$ | $4^3$ | $4^2$ |
| 5 | $1/3^4$ | $2^4$ | $2^3$ | $1/3^4$ | $4^4$ | $4^3$ |
| 6 | $1/3^5$ | $2^5$ | $2^4$ | $1/3^5$ | $4^5$ | $4^4$ |
| 7 | $1/3^6$ | $2^6$ | $2^5$ | $1/3^6$ | $4^6$ | $4^5$ |
| 8 | $1/3^7$ | $2^7$ | $2^6$ | $1/3^7$ | $4^7$ | $4^6$ |
| 9 | $1/3^8$ | $2^8$ | $2^7$ | $1/3^8$ | $4^8$ | $4^7$ |
| 10 | $1/3^9$ | $2^9$ | $2^8$ | $1/3^9$ | $4^9$ | $4^8$ |

Continued table 4

| Level | Sierpinski gasket | | | Vicsek snowflake | | |
|---|---|---|---|---|---|---|
| | Scale $L_m$ | Number $N_m$ | | Scale $L_m$ | Number $N_m$ | |
| | | Fractal | Complement | | Fractal | Complement |
| 1 | $1/2^0$ | $3^0$ | $(3^{-1})$ | $1/3^0$ | $5^0$ | $(4*5^{-1})$ |
| 2 | $1/2^1$ | $3^1$ | $3^0$ | $1/3^1$ | $5^1$ | $4*5^0$ |
| 3 | $1/2^2$ | $3^2$ | $3^1$ | $1/3^2$ | $5^2$ | $4*5^1$ |
| 4 | $1/2^3$ | $3^3$ | $3^2$ | $1/3^3$ | $5^3$ | $4*5^2$ |
| 5 | $1/2^4$ | $3^4$ | $3^3$ | $1/3^4$ | $5^4$ | $4*5^3$ |
| 6 | $1/2^5$ | $3^5$ | $3^4$ | $1/3^5$ | $5^5$ | $4*5^4$ |
| 7 | $1/2^6$ | $3^6$ | $3^5$ | $1/3^6$ | $5^6$ | $4*5^5$ |
| 8 | $1/2^7$ | $3^7$ | $3^6$ | $1/3^7$ | $5^7$ | $4*5^6$ |
| 9 | $1/2^8$ | $3^8$ | $3^7$ | $1/3^8$ | $5^8$ | $4*5^7$ |
| 10 | $1/2^9$ | $3^9$ | $3^8$ | $1/3^9$ | $5^9$ | $4*5^8$ |

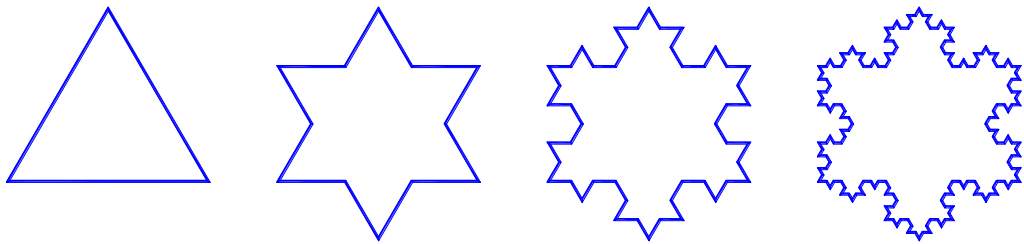

a. Koch snowflake curve

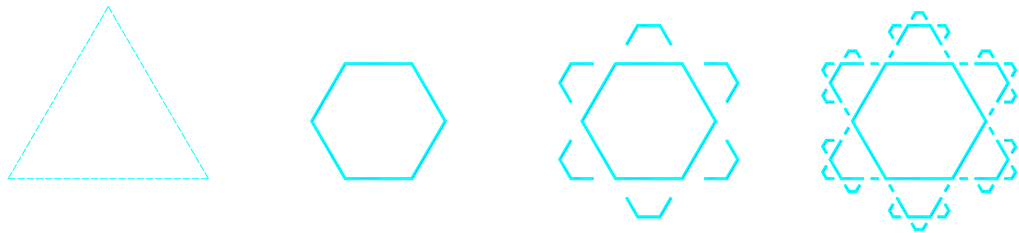

b. Koch complement

**Figure 10 Koch snowflake curve and its complementary set (the first four steps)**

[**Note**: The Koch snowflake can be employed to illustrate spatial development of the central place system dominated by market principle.]

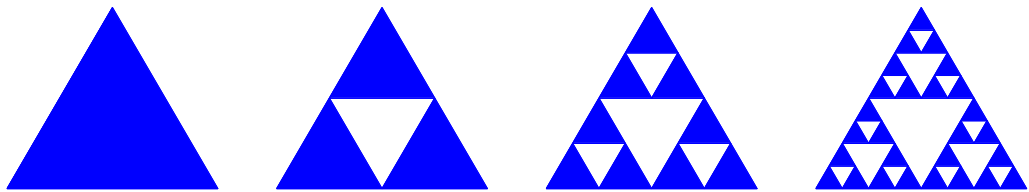

a. Fractal set

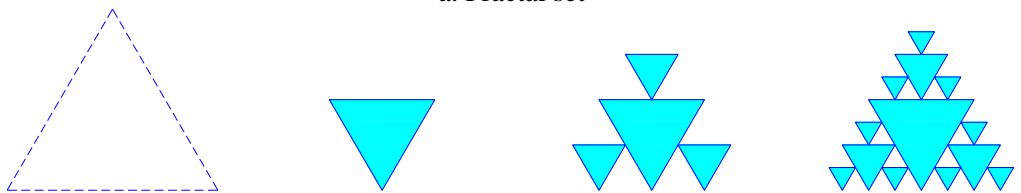

b. Fractal complement

**Figure 11 Sierpinski gasket and its complementary set (the first four steps)**

[**Note**: The Sierpinski curve can be employed to illustrate space filling of the central place system dominated by traffic principle.]

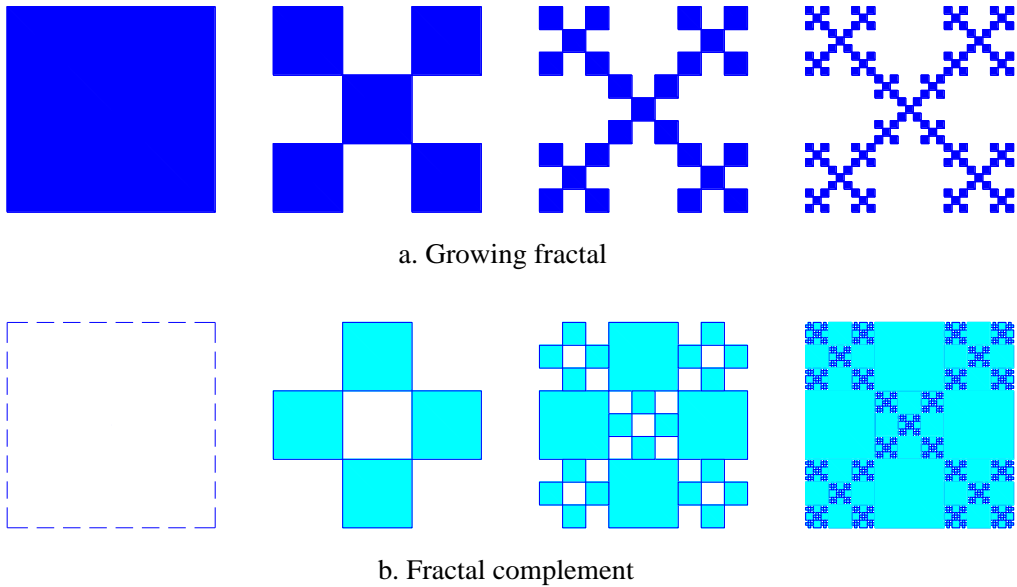

a. Growing fractal

b. Fractal complement

**Figure 12 Viscek snowflake and its complementary set (the first four steps)**

[**Note**: The Viscek snowflake bears an analogy with the growth and form of human settlements such as cities, towns, and villages.]

Studies on fractal complement hierarchies are useful in urban and rural geography. In many cases, special land uses such as vacant land, water areas, and green belts can be attributed to a fractal complement rather than a fractal set (Chen, 2010). However, this treatment is not necessary. Sometimes we specially evaluate the fractal parameter of vacant land, water areas, green belts, and so on. In particular, the spatial state of a settlement may be reversed: the fractal structure evolves into fractal complementary structure and *vice versa*. The concepts of fractals and fractal complements can be employed to model the evolution process of a settlement. If a fractal settlement is defined in a 2-dimensional space, its fractal dimension comes between 0 and 2 (Batty and Longley, 1994; Frankhauser, 1998; Feng and Chen, 2010). According to the spatial state and fractal dimension, the settlement evolution can be divided into four stages (Figure 13). The first stage is fractal growth (Figure 13(a)). In this stage, the geographical space is unstinted, and settlement growth bears a big degree of freedom. Typical phenomena are the new villages and young cities. The second stage is space filling (Figure 13(b)). In this stage, small fractal clusters appear in the vacant places. Typical phenomena are the mature cities, town, and villages. The third stage is structural reverse (Figure 13(c)). Settlement growth is a process of phase transition, which can be explained by space replacement dynamics. In this stage, the fractal structure of central part in the settlement is replaced by fractal complementary structure. The space dimension is near 2, which is a Euclidean dimension. Typical phenomena are the old cities, towns, and villages. Gradually, the central part becomes aging, degenerate, and finally has to be abandoned. Thus the settlements become hollow cities or hollow villages, from which inhabitants move away. The fourth stage is fractal regeneration (Figure 13(d)). After a period of desolation, the buffer space becomes large, and the central area is suitable for reconstruction. Thus, some people try to settle there in by rebuilding houses. In this stage, the fractal structure may become more complex and should be characterized by multifractal parameters.

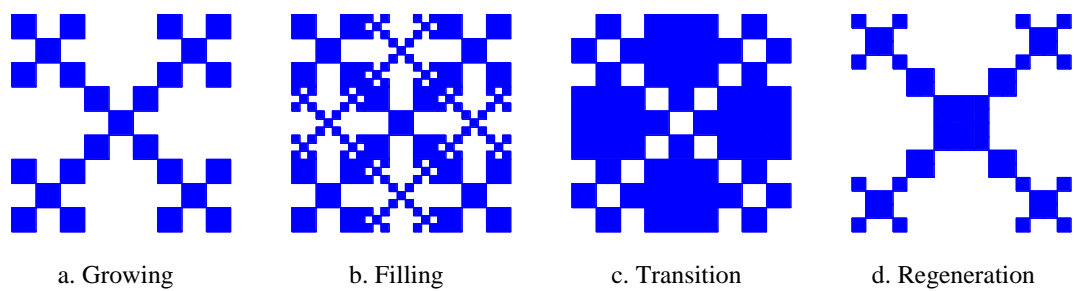

| a. Growing | b. Filling | c. Transition | d. Regeneration |

**Figure 13 A evolution model of human settlements with four development stages (the third step)**
[**Note**: Figure (a) is a simple fractal, figure (b) represents a large fractal growing together with four small fractals, figure (c) reflects a fractal complement, figure (d) indicates a multifractal growth. All the four patterns represent the third step of fractal growth.]

## 4.3 Logarithmic spiral and hierarchical scaling

The logarithmic spiral is also termed equiangular spiral or growth spiral, which is treated as a self-similar spiral curve in literature and is often associated with fractal such as the Mandelbrot set. The logarithmic spiral was first described by René Descartes in 1638 and later deeply researched by Jacob Bernoulli, who was so fascinated by the *marvelous spiral* that he wished it to be engraved on his tombstone. Hierarchical scaling can be employed to describe logarithmic spiral. Where geometric form is concerned, a logarithmic spiral bears an analogy with fractals; while where mathematical structure is concerned, the logarithmic spiral is similar to rank-size rule. Sometimes, the logarithmic spiral is treated as a fractal by scientists (Addison, 1997). In fact, a logarithmic spiral is not a real fractal because it has only one self-similarity point. For the section around the original point, the part of the logarithmic spiral is strictly similar to its whole. However, there is only self-similarity but there is no cross-similarity (Figure 14).

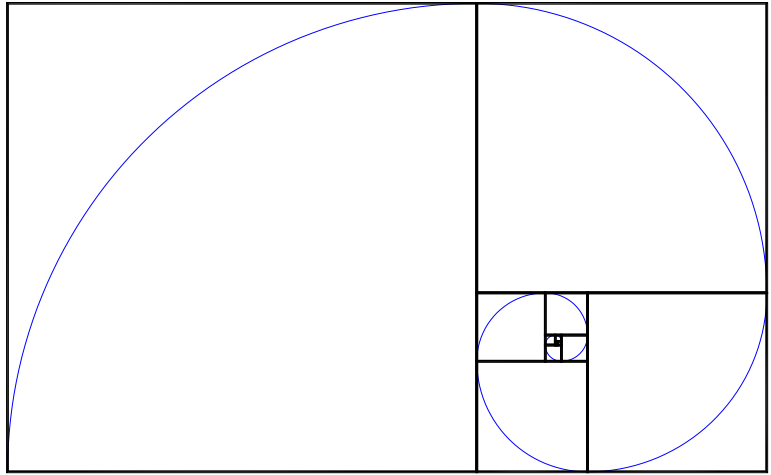

**Figure 14 A sketch map of spatial hierarchy for the logarithmic spiral based on the golden rectangle**

[**Note**: Using squares based on golden rectangles of different scales, we can derive a self-similar hierarchy for the logarithmic spiral, from which we can find an allometric scaling relation.]

Though a logarithmic spiral is not a fractal, this curve bears the similar mathematical model with simple fractals. A logarithmic spiral can be expressed as below:

$$x = a^\varphi \sin\theta = \sin\theta e^{(\ln a)\varphi} = \beta e^{\alpha\varphi}, \tag{36}$$

where $x$ denotes the distance from the origin, $\varphi$ is the angle from the abscissa axis, $\theta$ is a constant, and $\alpha = \ln(a)$ and $\beta = \sin(\theta)$ are two parameters. Integrating $x$ over $\varphi$ yields

$$L(\varphi) = \int_0^{m\pi} x(\varphi)\mathrm{d}\varphi = \beta\int_0^{m\pi} e^{\alpha\varphi}\mathrm{d}\varphi = \frac{\beta}{\alpha}\int_0^{m\pi} e^{\alpha\varphi}\mathrm{d}(\alpha\varphi) = \frac{\beta}{\alpha}[e^{\alpha m\pi} - 1] \to \infty, \tag{37}$$

where $L(\varphi)$ refers to a cumulative length. Thus we have

$$L_m(\varphi) = \beta\int_m^{m+1} e^{\alpha\varphi}\mathrm{d}\varphi = \frac{\beta}{\alpha}\left[e^{\alpha\varphi}\right]_m^{m+1} = \frac{\beta}{\alpha}(e^\alpha - 1)e^{\alpha m}, \tag{38}$$

in which $L_m(\varphi)$ denotes the length of the curve segment at the $m$th level. From equations (36) and (38), we can derive two common ratios

$$r_l = \frac{L_{m+1}(\varphi)}{L_m(\varphi)} = e^\alpha, \tag{39}$$

$$r_x = \frac{x_{m+1}}{x_m} = \frac{\beta e^{\alpha(m+1)}}{\beta e^{\alpha m}} = e^\alpha. \tag{40}$$

This suggest that the two common ratios are equal to one another, i.e., $r_l = r_x$. From equations (39) and (40) we can derive an allometric scaling relation such as

$$L_m = \kappa x_m^b, \tag{41}$$

where $\kappa$ refers to a proportionality coefficient, and $b$ to the scaling exponent. The allometric scaling relation indicates a special geometric measure relation. In fact, the allometric scaling exponent is

$$b = \frac{\ln r_l}{\ln r_x} = \frac{\alpha}{\alpha} = 1. \tag{42}$$

This result suggests a special allometric relation between two measurements of the logarithmic spiral. The above mathematical process shows that the logarithmic spiral as a quasi-fractal curve can be described strictly by hierarchical scaling.

In urban studies, the logarithmic spiral study is helpful for us to understand the central place theory about human settlement systems and the rank-size distribution of cities. Central place systems are composed of triangular lattice of points and regular hexagon area (Christaller, 1933/66). From the regular hexagonal networks, we can derive logarithmic spiral (Thompson, 1966). On the other hand, the mathematical models of hierarchical structure of the logarithmic spiral based on the systems of golden rectangles are similar to the models of urban hierarchies based on the rank-size distribution. The logarithmic spiral suggests a latent link between Zipf's law indicating hierarchical structure and Christaller's central place models indicative of both spatial and hierarchical structure. Maybe we can find new spatial analytical approach or spatial optimization theory by exploring the hierarchical scaling in the logarithmic spiral.

## 5 Conclusions

The conventional mathematical modeling is based on the idea of characteristic scales. If and only if a characteristic length is found in a system, the system can be effectively described with traditional mathematical methods. However, complex systems are principally scale-free systems, and it is hard to find characteristic lengths from a complex system. Thus mathematical modeling is often ineffectual. Fractal geometry provides a powerful tool for scaling analysis, which can be applied to exploring complexity associated with *time lag*, *spatial dimension*, and *interaction*. However, any scientific method has its limitation. Fractal description bears its sphere of application. In order to strengthen the function of fractal analysis, hierarchical scaling theory should be developed. Fractal analytical process can be integrated into hierarchical scaling analysis. In this work, three aspects of studies are presented. *First, hierarchical scaling is a simple approach to describing fractal structure*. Fractal scaling used to be expressed with power laws. Based on hierarchical structure, a power law can be transformed into a pair of exponential laws, and the analytical process is significantly simplified. *Second, fractal analysis can be generalized to quasi-fractal phenomena such as logarithmic spiral*. A real fractal possesses more than self-similarity point, while logarithmic spiral has only one self-similarity point. Using hierarchical scaling, fractals and quasi-fractals can be modeled in its right perspective. *Third, spatial analysis can be associated with hierarchical analysis*. Spatial dimension is one of obstacles for mathematical modeling and analysis. It is more difficult to make spatial analysis than hierarchical analysis. By hierarchical scaling, a spatial network can be transformed into a hierarchy with cascade structure, and the spatial analysis can be equivalently replaced by hierarchical analysis. According to the above-mentioned ideas, we can develop an integrated theory based on fractal and hierarchical scaling to research complex systems such as cities. What is more, fractals reflect optimum structure in nature. A fractal object can occupy its space in the most efficient way. Using concepts from fractals and hierarchical scaling, we can optimize human settlement systems, including cities, towns, villages, and systems of cities and towns.

## Acknowledgements

This research belongs to the Key Technology R&D Program of the National Ministry of Science and Technology of China (Grant No. 2014BAL01B02). The financial support is gratefully acknowledged.